# COSMOLOGICAL VELOCITY BIAS

R. G. CARLBERG

*Department of Astronomy, University of Toronto, Toronto, Canada M5S 1A1*

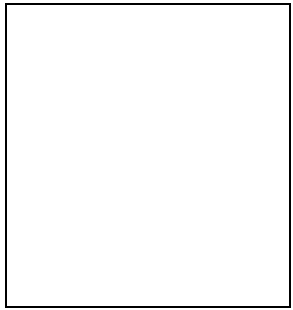

**Abstract**

Velocity bias is a reduction of the velocity dispersion of tracer galaxies in comparison to the velocity dispersion of the underlying mass field. There are two distinct forms of velocity bias. The single particle velocity reduction, $b_v(1)$, is the result of energy loss of a tracer population, and in virialized regions, such as galaxy clusters, is intimately associated with mass segregation which together lead to cluster mass underestimates. The pairwise velocity bias, $b_v(2)$, has an additional statistical reduction if the total mass per galaxy rises with velocity dispersion of the virialized cluster. Values of the velocity bias are estimated from n-body simulations, finding $b_v(1) \simeq 0.85 \pm 0.1$ and $b_v(2) \simeq 0.6 \pm 0.2$. The value of $b_v(1)$ is relatively secure and predicts that the virial radius of the cluster light is about 20% of the cluster mass, which can be tested with observations of cluster mass profiles beyond the apparent virial radius. The value of $b_v(2)$ is sensitive to the formation efficiency of galaxies over environments ranging from voids to rich clusters, the latter of which are not yet well resolved in simulations. An $n=1$, $\Omega=1$, COBE normalized CDM spectrum requires $b_v(2) \simeq 0.20(15\mu K/Q)(\sigma_{12}/317\,{\rm km\,s^{-1}})$ which is well below the measured range of $b_v(2)$. The pairwise velocities at $1\,h^{-1}$ Mpc allow $\Omega=1$ for a galaxy clustering bias near unity if $b_v(2) \simeq 0.6$.

## 1 Introduction

There are compelling theoretical arguments favouring an $\Omega=1$ universe [21, 1] but the pairwise random velocities predicted from the observed galaxy clustering are about twice those that are observed in the galaxy population [14, 15]. Furthermore studies of large clusters of galaxies generally find an implied $\Omega \simeq 0.2$ [28]. These observations do not conclusively rule out $\Omega=1$, since it is not known whether or not galaxies accurately represent the dynamics of the mass field. For instance, bulk streaming velocities, measured on scales where the clustering is nearly linear, indicate that $\Omega^{0.6}/b \simeq 1$ (with substantial error bars, and considerable uncertainty as to the value of $b$) [4, 25, 33]. The "small" small scale velocities are utterly devastating for the CDM spectrum [7] when normalized with the COBE detection of cosmic microwave background fluctuations [34, 35], predicting $\Omega=1$ pairwise random velocities nearly a factor of 5 higher than observed. The discrepancy between the small and large scale $\Omega$ values can be understood if a velocity bias affecting the random velocities of clustered galaxies exists. The primary aim of this paper is to provide current n-body estimates of the degree of velocity bias that might be present.

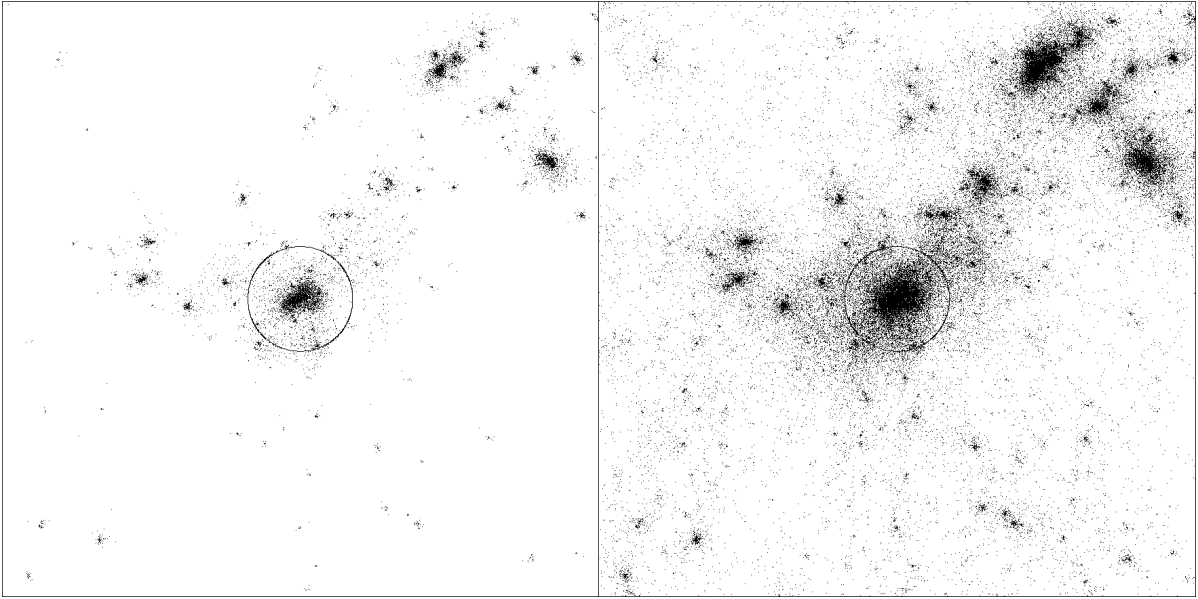

Figure 1: The distribution of tracer particles (left) and all particles (right) in the vicinity of a large cluster plotted within a box with edges of 10 $h^{-1}$ Mpc. The circle is the virial radius of the mass distribution. The tracer particles were identified as all particles at $z=3$ located in groups of at least 5 particles linked together with a separation less than 0.1 of the mean separation. The velocity bias of the ungrouped tracer particles is $b_v(1) \simeq 0.85$.

Velocity bias has been reported in a number of cosmological simulations that have hierarchical galaxy formation [9, 10, 11, 8, 12, 13, 18, 26, 20, 27] but disagree in detail on the size of the effect and its causes. Velocity bias exists only in the strongly nonlinear regime—infall velocities are close to their expected values [8]. The observation of velocity bias in simulations is partly an issue of how galaxies are identified from the particle distribution. There is not yet an agreed upon identification procedure in principle, and for any procedure at current numerical resolutions there are problems in practice. The approach to galaxy identification here is that galaxies only form in the dense central regions of dark matter halos having velocity dispersions in the range around 100 km s$^{-1}$ [38] and then, once formed, galaxies can merge with other galaxies. This paper reports the results of simulations having sufficient particles and length resolution that the dense cores of halos contain the same particles over a cluster dynamical time, thereby diminishing the parameter dependence of the galaxy selection algorithm. The problem of two body relaxation gradually destroying dense halos in clusters is serious, and will not be solved until much larger simulations are available.

## 2  Single Particle Velocity Bias

Velocity bias, $b_v \equiv \sigma_t/\sigma_\rho$, has two distinct forms. The single particle velocity bias, $b_v(1)$, is a cooling in virialized clusters of tracer population velocities, $\sigma_t$, as compared to the mass field, $\sigma_\rho$, [11]. The pairwise velocity bias, $b_v(2)$, is an average over the field and can have a statistical reduction in addition to the velocity cooling [5]. The virial analysis of cluster mass to light ratios is affected by $b_v(1)$. The pairwise bias $b_v(2)$ appears in the cosmological virial theorem [29].

Single particle velocity bias gives rise to a segregation of the tracer population with respect to the background mass field. A cool, light, tracer population in an equilibrium system with a power law mass density profile, $\rho(r) = b_\rho r^{-n}$, is in general more centrally concentrated than the mass distribution. For a spherical system with an isotropic velocity ellipsoid (the result can be generalized) the self-gravitating "background" mass has a velocity dispersion given by the Jeans equation, $d\rho\sigma^2/dr = -\rho\nabla\phi$, to be,

$$\sigma_b^2(r) \equiv b_\sigma r^{2-n} = \frac{4\pi G b_\rho}{2(n-1)(3-n)} r^{2-n}. \qquad (1)$$

The range of validity is $1 < n < 3$, which is the local density gradient for all but the inner and outermost part of a typical dark halo [17]. A cool tracer population with velocity dispersion $\sigma_c^2(r) = c_\sigma r^{-q}$ and density profile $\rho_c(r) = c_\rho r^{-p}$ inserted into this background potential must satisfy the Jeans equation,

$$(p+q)c_\sigma r^{-q-1} = 2(n-1)b_\sigma r^{-n+1}, \qquad (2)$$

from which we find, $q = n - 2$, the same radial dependence of the velocity dispersion as for the background population. Noting that $c_\sigma/b_\sigma = b_v^2(1)$,

$$p = n + 2(n-1)\left(\frac{1}{b_v^2(1)} - 1\right). \qquad (3)$$

Hence a cool background population, $b_v(1) < 1$, will have $p > n$. That is, the tracer will be a steeper, more centrally concentrated distribution. Because the exponent of the power law depends on the ratio of velocity dispersions, a small temperature difference translates into a large radial segregation. The background mass interior to radius $r$ is $M(r) = 4\pi b_\rho/(3-n)r^{3-n}$, with a similar relation for the cool population. If both masses are normalized at some common outer radius, half the turnaround radius for instance, then the ratio of the half mass radii is,

$$\frac{r_h^c}{r_h^b} = \left(\frac{1}{2}\right)^{\frac{1}{3-p} - \frac{1}{3-n}}. \qquad (4)$$

If the background profile is taken as "isothermal", $n = 2$, and the cool population has a velocity dispersion $\sigma_c = \sqrt{4/5}\sigma_b = 0.894\sigma_b$, giving $p = 5/2$, then $r_h^c/r_h^b = 1/2$, a dramatic effect. A detailed analysis using Hernquist's potential [23], which is analytically well behaved at all radii, gives essentially identical results for the half mass radius and shows that the apparent virial radius of a cool population in equilibrium drops by a factor of 6 for $b_v(1) = 0.9$. Figure 1 illustrates (see also [36, 11]) the strong mass segregation implied by this small single particle velocity bias.

## 3 Substructure Dynamics in a Cluster

The physical cause of single particle velocity bias is understood in principle, but not in quantitative detail, as an energy loss of the dense core through dynamical friction on the cluster material [36, 8], and asymmetric ejection of the low density envelope [22]. The visible, virialized galaxies in large clusters appear to be essentially free of dynamical friction [37]. However, infalling galaxies are the dense, visible cores of massive correlated halos several hundred times heavier which are sufficient for considerable dynamical friction.

Identifying the sites of galaxies in a simulation is a fundamental problem for the application of simulation results to the observable universe. Structure formation via hierarchical gravitational instability suggests a straightforward galaxy identification procedure: locate the dense central regions of halos with dispersion in the range around 100 km s$^{-1}$ [32, 36, 6, 39]. Ultimately galaxy identification will improve with gas and stars added to simulations, however at the moment all the computing power available marginally resolves galaxy scales in a cluster simulation. Since two-body heating of cool substructures in hot backgrounds is a severe problem in particle simulations the approach here is to do an entirely gravitational simulation, and to demonstrate that the dense dark matter cores persist in the virialized cluster.

The xy projection of the final state of the million particle cluster simulation is shown in Figure 2. The important new feature is the presence of a substantial amount of low mass substructure that has survived a number of orbits within this well virialized cluster. This substructure is naturally identified as galaxy halos. The innermost region of the cluster has a deficiency of substructure as a result of two body evaporative heating. The initial conditions for the simulation were created using the CDM spectrum [7] realized on a $128^3$ grid in a box $50h^{-1}$ Mpc on a side. The position of the largest peak in the box was located by smoothing with a $2.5h^{-1}$ Mpc Gaussian filter. A sphere of